\documentclass[]{aa}
\usepackage{graphicx}
\usepackage{color}
\usepackage{txfonts}

\begin{document}

\title{Cut-off wavenumber of Alfv\'en waves in partially ionized plasmas of the solar atmosphere}

\author{Zaqarashvili, T. V.\inst{1,4}, Carbonell, M. \inst{2}, Ballester J. L. \inst{3} and Khodachenko, M. L. \inst{1}
}

 \institute{Space Research Institute, Austrian Academy of Sciences, Schmiedlstrasse 6, 8042 Graz, Austria\\
             \email{[teimuraz.zaqarashvili;maxim.khodachenko]@oeaw.ac.at}
                               \and
 Departament de Matem\`{a}tiques i Inform\`{a}tica. Universitat de les Illes Balears, E-07122 Palma de Mallorca, Spain\\
 \email{marc.carbonell@uib.es}
 \and
 Departament de F\'{\i}sica, Universitat de les Illes Balears, E-07122 Palma de Mallorca, Spain\\
\email{joseluis.ballester@uib.es}
\and
            Abastumani Astrophysical Observatory at Ilia State University, University St. 2, Tbilisi, Georgia\\
}

\date{Received / Accepted }

\abstract{Alfv\'en wave dynamics in partially ionized plasmas of the solar atmosphere shows that there is indeed a cut-off wavenumber, i.e. the Alfv\'en waves with wavenumbers higher than the cut-off value are evanescent. The cut-off wavenumber appears in single-fluid magnetohydrodynamic (MHD)   approximation but it is absent in a multi-fluid approach. Up to now, an explanation for the existence of the cut-off wavenumber is still missing.}{The aim of this paper is to point out the reason for the appearance of a cut-off wavenumber in single-fluid MHD.}{Beginning with three-fluid equations (with electrons, protons and neutral hydrogen atoms), we performed consecutive approximations until we obtained the usual single-fluid description is obtained. We solved the dispersion relation of linear Alfv\'en waves at each step and sought the approximation responsible of the cut-off wavenumber appearance.}{We have found that neglecting inertial terms significantly reduces the real part of the Alfv\'en frequency although it never becomes zero. Therefore, the cut-off wavenumber does not exist at this stage. However, when the inertial terms together with the Hall term in the induction equation are neglected, the real part of the Alfv\'en frequency becomes zero.}{The appearance of a cut-off wavenumber, when Alfv\'en waves in partially ionized regions of the solar atmosphere are studied, is the result of neglecting  inertial and Hall terms, therefore it has no physical origin.}


\titlerunning{Cut-off wavenumber of Alfv\'en waves}

\authorrunning{Zaqarashvili et al.}

\maketitle

\section{Introduction} \label{intro}

Cut-off wavenumbers in fully ionized resistive MHD have been reported in several classical textbooks. For instance in the chapter on the study of the thermal instability of a layer of fluid heated from below when a magnetic field is present, Chandrasekhar (\cite{Chandrasekhar1961}) described the behavior of Alfv\'en waves in a viscous and resistive medium. In this case, the Alfv\'en wave frequency, $\omega$, is given by
\begin{equation}
\omega = \pm k\sqrt{V_A^2-\frac{1}{4}(\nu - \eta)^2 k^2}-\frac{1}{2}i(\nu+\eta)k^2, \label{cw0}
\end{equation}
where $V_A$ is the Alfv\'en speed, $\nu$ the kinematic viscosity, $\eta$ the magnetic diffusivity, and $k$ the wavenumber. This expression clearly points out that for a value of the wavenumber such as
\begin{equation}
k = \pm \frac{2V_A}{\nu - \eta} \label{cw}
\end{equation}
the real part of the frequency becomes zero, and only the imaginary part of the frequency remains. The cut-off wavenumber means that waves with a wavenumber higher than the cut-off value are evanescent. Therefore, Eq. (\ref{cw}) provides us with the cut-off wavenumber for resistive and viscous plasmas when the single-fluid approximation is used. When viscosity is negligible, but resistivity still exists (resistive plasma), this cut-off wavenumber is modified by keeping only the resistivity in the denominator of Eq. (\ref{cw}). Chandrasekhar (\cite{Chandrasekhar1961}) made no explicit comment about this cut-off wavenumber, only pointed out that when viscosity or resistivity are present, Alfv\'en waves are damped. Ferraro \& Plumpton (\cite{Ferraro1961}) and Kendall \& Plumpton (\cite{Kendall1964}) considered the effects of finite conductivity on hydromagnetic waves, and showed that for $\eta k < 2V_A$, we have time-damped waves, while for $\eta k > 2V_A$ there is no wave propagation at all. Furthermore, Cramer (\cite{Cramer2001}) has pointed out the same effect, showing that when the wavenumber becomes greater than $2 R_m/L$, where $R_m$ is the magnetic Reynolds number and $L$ a reference length, the real part of the Alfv\'en wave frequency becomes zero. Although some of the above reported textbooks have shown that there is a cut-off wavenumber, none of them has investigated the reason for its presence in resistive single-fluid MHD and its physical meaning, if any.

On the other hand, the real part of Eq.(\ref{cw0}) can be written as
\begin{equation}
\omega_r = \pm k \Gamma_A \label{cw1},
\end{equation}
with
\begin{equation}
\Gamma_A = V_A \sqrt{1-\frac{1}{4 V_A^2}(\nu - \eta)^2 k^2} \label{cw2}
\end{equation}
representing a modified Alfv\'en speed that goes to zero for the cut-off wavenumber, i. e. the wave ceases its propagation.

Significant parts of the solar atmosphere, namely photosphere, chromosphere and prominences are only partially ionized. Collisions between different plasma species generally lead to the damping of waves in magnetized plasmas. Neutral atoms may enhance the damping of transverse waves through ion-neutral collision (Braginskii \cite{Braginskii1965}, Khodachenko el al. \cite{Khodachenko2004}, Carbonell et al. \cite{Carbonell2010}, Zaqarashvili et al. 2011a,b) and may lead to heating of the ambient plasma (Khomenko and Collados \cite{Khomenko2012}). In partially ionized plasmas and in an astrophysical context, Kulsrud \& Pearce (\cite{Kulsrud1969}) studied the interaction between cosmic rays and Alfv\'en waves in the interstellar medium, and showed that a cut-off wavenumber appeared there as well. Balsara (\cite{Balsara1996}) stu\-died MHD wave propagation in molecular clouds in the single-fluid approximation, reporting that Alfv\'en and fast waves become completely damped when the wavenumber attains a certain value. Recently, and in the context of solar prominence oscillations, cut-off wavenumbers have been reported again. Forteza et al. (\cite{Forteza2007}, \cite{Forteza2008}), Barcel\'o et al. (\cite{Barcelo2011}) and Soler et al. (\cite{Soler2011}) used the single-fluid approximation to study the damping of MHD waves produced by ion-neutral collisions in an unbounded medium with prominence physical pro\-perties. These authors found that the cut-off wavenumber for Alfv\'en waves is given by
\begin{equation}
k = \pm \frac{2V_A \cos \theta}{(\eta_c \cos^2 \theta + \eta \sin^2 \theta)} \label{ck},
\end{equation}
with $\theta$ the propagation angle with respect to the magnetic field, and $\eta_c$ the Cowling diffusivity. In this case, the modified Alfv\'en speed (Barcel\'o et al. 2011) is given by
\begin{equation}
\Gamma_A = V_A \sqrt{1-\frac{(\eta_c \cos^2 \theta +\eta \sin^2 \theta)^2 k^2}{4 V_A^2 \cos^2 \theta}}. \label{cw3}
\end{equation}
For fully ionized resistive plasmas, $\eta = \eta_c$ and, for parallel propagation, we recover the Kendall \& Plumpton (\cite{Kendall1964}) cut-off wavenumber. When fast and slow waves are coupled, the nume\-rical value of the cut-off wavenumber for fast waves is similar to that of Alfv\'en waves, but for parallel propagation fast waves decouple, becoming Alfv\'en waves, and their cut-off wavenumber is exactly the same as for Alfv\'en waves. Singh \& Krishnan (\cite{Singh2010}) studied the behavior of Alfv\'en waves in the partially ionized solar atmosphere and they also reported the cut-off wavenumber.

Since time-damped transversal oscillations have been observed in the fine structure of solar filaments,  Soler et al. (\cite{Soler2009a}, \cite{Soler2009b}) modeled this fine structure as magnetic cylinders filled with partially ionized plasma, which has prominence physical properties, and studied the time damping of transverse oscillations produced by ion-neutral collisions and resonant absorption. Again, cut-off wavenumbers for fast and Alfv\'en waves appear, as described by Eq.(\ref{ck}).

The behaviour of the cut-off wavenumber in a partially ionised plasma can be understood by writing its expression in terms of $\eta$ and $\eta_C$.
The Cowling magnetic diffusivity, $\eta_C$, is
\begin{eqnarray}
\eta_C = \frac{1}{4 \pi}\left[\frac{m_e c^2}{n_e e^2}\left (\frac{1}{\tau_{ei}}+\frac{1}{\tau_{en}}\right )+\frac{\xi_n^2 B_0^2}{\alpha_n}\right],
\end{eqnarray}
where $\tau_{en}$ and $\tau_{ei}$ are the electron-neutral and electron-ion collisional times, respectively, $\xi_{n}$ represents the relative density of neutrals, and $\alpha_n$ is a friction coefficient whose expression is
\begin{eqnarray}
\alpha_n = \frac{1}{2}\xi_n(1-\xi_n)\frac{\rho_0^2}{m_i}\sqrt{\frac{16k_BT_0}{\pi m_i}}\Sigma_{in},
\end{eqnarray}
with $\Sigma_{in}$ the ion-neutral collision cross-section. The Spitzer magnetic diffusivity, $\eta$, is
\begin{eqnarray}
\eta = \frac{c^2}{4 \pi} \frac{m_e }{n_e e^2}\left [\frac{1}{\tau_{ei}}+\frac{1}{\tau_{en}}\right ],
\end{eqnarray}
then, the cut-off wavenumber, $k$, given by Eq.~(\ref{ck}), becomes
\begin{equation}
 k= \frac{2  V_A \cos \theta}{\frac{1}{4 \pi} \Big[\frac{m_e c^2}{n_e e^2}\Big(\frac{1}{\tau_{ei}}+\frac{1}{\tau_{en}}\Big)+\frac{\xi_n^2 B_0^2}{\alpha_n}\cos^2 \theta  \Big]
}.
 \end{equation}


 On the other hand, as reported by Cramer (\cite{Cramer2001}), the cut-off wavenumber, $k$, can be written in terms of the magnetic Reynolds number. In a partially ionized plasma of the solar atmosphere, the Cowling magnetic diffusivity, $\eta_C$, is several orders of magnitude greater than Spitzer's magnetic diffusivity, $\eta$, therefore, neglecting $\eta$ in Eq.(\ref{ck}), we obtain
\begin{eqnarray}
k = \pm \frac{2 V_A}{\eta_C \cos \theta} \label{kc}.
\end{eqnarray}
In our case,  the magnetic Reynolds number, $R_\mathrm{m}$, could be written as $R_\mathrm{m} = uL/\eta_C$, where $u$ is a characteristic velocity, which we could take as the Alfv\'en speed, and $L$ is a characteristic scale-length. Then, $R_\mathrm{m}$ can be expressed as $R_\mathrm{m} = V_A L/\eta_C$, and using Eq.~(\ref{kc}) we obtain
\begin{eqnarray}
k= \frac{2 R_\mathrm{m}}{L \cos \theta}. \label{kr}
\end{eqnarray}

Summarizing, cut-off wavenumbers appear when the single-fluid approximation is used to study MHD waves in collisional magnetized astrophysical plasmas. In particular, this topic is relevant in connection with MHD waves in regions of the solar atmosphere such as chromosphere and photosphere, or in chromospheric and coronal solar structures such as spicules or prominences. However, and up to now, an explanation for the existence of the cut-off wavenumber is still missing.

Recently, Zaqarashvili et al. (\cite{Zaqarashvili2011a}) have shown that the cut-off wavenumber for Alfv\'en waves is absent in a two-fluid MHD approximation, where one fluid is the ion-electron gas and the other fluid is the gas of neutral hydrogen. Then, it is clear that the cut-off wavenumber appears as a result of an approximation and is not an intrinsic property of waves.

Here we study the dynamics of linear Alfv\'en waves in the partially ionized plasma of the solar atmosphere and seek for the reason for the appearance of the cut-off wavenumber. We start from the full multi-fluid linear equations and make consecutive approximations until the usual equations for Alfv\'en waves in single-fluid MHD are recovered. After each approximation, we derive the corresponding dispersion relations which, once solved, allow us to uncover the approximation responsible for the cut-off wavenumber appearance.


\section{Main equations}

We study partially ionized plasmas made of
electrons (${\mathrm {e}}$), ions (${\mathrm {i}}$) and neutral (hydrogen) atoms (${\mathrm {n}}$). We suppose that each sort of
species possess Maxwell velocity distribution, therefore they can be described as separate fluids.

We aim to study the dynamics of Alfv\'en waves, therefore we consider incompressible plasma. We also neglect the viscosity, the heat flux and the heat production due to collisions between particles. Plasma and unperturbed magnetic field are considered to be homogeneous. Then, linearized fluid equations for each species can be split into perpendicular and parallel components of perturbations with regard to the unperturbed magnetic field. We are interested in Alfv\'en waves, therefore only perpendicular components are considered. Then, the fluid and Maxwell (without displacement current) equations can be written as

\begin{equation}\label{ne3}
\nabla \cdot \vec u_{a\perp}=0,
\end{equation}
\begin{equation}\label{Ve3}
m_an_a{{\partial \vec u_{a\perp}}\over {\partial t}}=-e_a n_a\left (\vec E_{\perp} +{1\over c}\vec u_{a\perp} \times \vec B \right )+\vec
R_{a\perp},
\end{equation}
\begin{equation}\label{e}
\nabla \times \vec E_{\perp}=-{1\over c}{{\partial \vec b_{\perp}}\over {\partial t}},
\end{equation}
\begin{equation}\label{B}
\nabla \times \vec b_{\perp}={{4 \pi} \over c}\vec j_{\perp},\,\,\, \vec j_{\perp}=-en_{\mathrm {e}}(\vec u_{\mathrm {e\perp}}-\vec u_{{\mathrm {i\perp}}}),
\end{equation}
where $m_a$ and $n_a$ are the mass and the number
density of particles $a$, $\vec B$ is the unperturbed magnetic field strength, $\vec u_{a\perp}$ and $\vec b_{\perp}$ are the velocity and magnetic perturbations perpendicular to the unperturbed magnetic field, $\vec E_{\perp}$ and $\vec j_{\perp}$ are the perpendicular components of electric field and current density, $e_a=\pm 4.8\times 10^{-10}$ statcoul is the charge of electrons and protons (note, that $e_a=0$ for neutral hydrogen) and $c=2.9979\times 10^{10}$ cm s$^{-1}$ is the speed of light. Plasma is supposed to be quasi-neutral, which means
$n_{\mathrm {e}}=n_{\mathrm {i}}$. $\vec R_{a\perp}$ is the change of impulse of particles $a$ due to
collisions with other sort of particles, which in the case of Maxwell distribution in each sort of particles has the form (Braginskii \cite{Braginskii1965})
\begin{equation}\label{Re}
\vec R_{a\perp}=-\sum_b \alpha_{ab}(\vec u_{a\perp}-\vec u_{b\perp}),
\end{equation}
where $\alpha_{ab}=\alpha_{ba}$ are coefficients of friction between particles $a$ and $b$.

For time scales longer than ion-electron and ion-ion collision times, the
electron and ion gases can be considered as a single fluid. This
significantly simplifies the equations taking into account the
smallness of electron mass with regard to the masses of ion and
neutral atoms. Then the three-fluid description can be changed by
a two-fluid description, where one component is the charged fluid (electron+protons) and the other component is the gas of neutral hydrogen (Zaqarashvili et al. 2011a).

Then we may go a step further and derive the single-fluid MHD equations. We use the total velocity (i.e. velocity of center of mass)
\begin{equation}
\vec u_{\perp}= {{\rho_{i}\vec u_{i\perp}+\rho_{n}\vec u_{n\perp}}\over
{\rho_{i}+\rho_{n}}},\,\
\end{equation}
the relative velocity
\begin{equation}\label{w}
\vec w_{\perp}= \vec u_{i\perp} - \vec u_{n\perp}.
\end{equation}
and the total density
\begin{equation}\label{rho}
\rho=\rho_{i}+\rho_{n},
\end{equation}
where $\rho_a=m_a n_a$ is the density of the corresponding fluid. Then, Eqs. (\ref{ne3}-\ref{B}) are rewritten as
\begin{equation}\label{V1}
\rho{{\partial \vec u_{\perp}}\over {\partial t}}={1\over {4\pi}}(\nabla \times \vec b_{\perp}) \times \vec B,
\end{equation}
\begin{equation}\label{w1}
{{\partial \vec w_{\perp}}\over {\partial t}}={1\over {4\pi
\rho \xi_i}}(\nabla \times \vec b_{\perp}) \times \vec B +{{c \alpha_{en} }\over {4\pi
e n_e \rho \xi_i \xi_n}}\nabla \times \vec b_{\perp}-{{\alpha_{in}+\alpha_{en}}\over {\rho \xi_i \xi_n}}\vec w_{\perp},
\end{equation}
$$
{{\partial \vec b_{\perp}}\over {\partial t}}={\nabla \times}(\vec u_{\perp} \times
\vec B)-\eta {\nabla \times}(\nabla \times \vec b_{\perp})-{{c}\over {4\pi en_e}} {\nabla
\times}\left((\nabla \times \vec b_{\perp})\times \vec B \right )+
$$
\begin{equation}\label{B1}
+{{c \alpha_{en}}\over {en_e}} {\nabla
\times}\vec w_{\perp}+\xi_n{\nabla \times}\left (\vec w_{\perp} \times \vec B\right ),
\end{equation}
where $\xi_i=\rho_i/\rho$, $\xi_n=\rho_n/\rho$  and
\begin{equation}\label{etaT}
\eta={{c^2}\over {4\pi \sigma}}={{c^2}\over {4\pi e^2 n^2_e}}\left [\alpha_{ei}+{{\alpha_{ei}\alpha_{en}}\over {\alpha_{ei}+\alpha_{en}}} \right ].
\end{equation}


The single-fluid Hall MHD equations are obtained from Eqs. (\ref{V1}-\ref{B1}) as follows.
The inertial term (the left-hand side term in Eq. \ref{w1}) is neglected, which is a good approximation for time scales longer than ion-neutral collision time, but fails for the shorter time scales (Zaqarashvili et al. 2011a). Then $\vec w_{\perp}$ defined from Eq. (\ref{w1}) is substituted into Eq. (\ref{B1}) and one can obtain the Hall MHD equations
\begin{equation}\label{V2}
\rho{{\partial \vec u_{\perp}}\over {\partial t}}={1\over {4\pi}}(\nabla \times \vec b_{\perp}) \times \vec B,
\end{equation}
$$
{{\partial \vec b_{\perp}}\over {\partial t}}={\nabla \times}(\vec u_{\perp} \times
\vec B)-{{c}\over {4\pi en_e}} \Big [1- {{2\xi_n\alpha_{en}}\over {\alpha_{in}+\alpha_{en}}}\Big ]{\nabla
\times}\left((\nabla \times \vec b_{\perp})\times \vec B \right )+
$$
\begin{equation}\label{B2}
+\eta_c \nabla^2 \vec b_{\perp},
\end{equation}
where
\begin{equation}\label{etac}
\eta_c=\eta + {{\xi^2_n B^2}\over {4\pi \alpha_{in}}},
\end{equation}
is the Cowling coefficient of magnetic diffusion and the second term in the right side of Eq. (\ref{B2}) is the Hall current term modified by electron-neutral collisions.

The usual single-fluid MHD equations, which are widely used for describing of Alfv\'en waves in partially ionized plasmas, are obtained from Eqs. (\ref{V2}-\ref{B2}) after neglecting the modified Hall term in Eq. (\ref{B2})
\begin{equation}\label{V3}
\rho{{\partial \vec u_{\perp}}\over {\partial t}}={1\over {4\pi}}(\nabla \times \vec b_{\perp}) \times \vec B,
\end{equation}
\begin{equation}\label{B3}
{{\partial \vec b_{\perp}}\over {\partial t}}={\nabla \times}(\vec u_{\perp} \times
\vec B)+\eta_c \nabla^2 \vec b_{\perp}.
\end{equation}

\section{Alfv\'en waves in partially ionized plasmas of the solar atmosphere}

We study the propagation of linear Alfv\'en waves in partially ionized plasmas of the solar atmosphere. Here, and for the numerical calculations, we have chosen physical parameters corresponding to solar quiescent prominences ($B=10$ G, $n_e=10^{10}$ cm$^{-3}$, $T=8000$ K). The unperturbed magnetic field $B$ is directed along the $z$ axis of cartesian frame. Next, we consider the Alfv\'en wave propagation along the magnetic field, consequently, we perform the Fourier analysis with $exp(-i\varpi t + ik z)$, where $\varpi$ is the wave frequency and $k$ is the wavenumber.

From Eqs. (\ref{V1}-\ref{B1}) the following dispersion relation is obtained
$$
a\delta^2 \nu \left [1+(1+\nu)\zeta \right ]\omega -a \xi_n \left [(\pm a + \xi_i \omega)\omega -1 \right ]\omega +i \delta \Big [a^2 \xi_n \omega^2+
$$
$$
+ \nu\Big [\pm a \omega (\zeta-1)-a^2 \zeta \omega^2 +\xi_i (1+\zeta(1\mp 2a \omega)+((a^2-1)\zeta-1)\omega^2)\Big ]\Big ]=
$$
\begin{equation}\label{dispn1}
=0,
\end{equation}
where $\omega=\varpi/(k V_A)$, $\tau=\omega_e/\omega_i$, $a={{k V_A}/ {\omega_i}}$, $\delta=\delta_{ei}/\omega_e$, $\nu=\alpha_{in}/\alpha_{ei}$ and $\zeta=\alpha_{en}/\alpha_{in}$. Here $V_A=B/\sqrt{4 \pi \rho}$ is the Alfv\'en speed, $\delta_{ei}=\alpha_{ei}/(m_e n_e)$ is the electron-ion collision frequency, $\omega_i=e B/(c m_i)$ and $\omega_e=e B/(c m_e)$ are ion and electron giro-frequencies, respectively.

%

The dispersion relation given by Eq. (\ref{dispn1}) has six different solutions. Four of the solutions are Alfv\'en waves polarized in two perpendicular
planes and propagating in opposite directions. The remaining solutions have zero real frequency at low wavenumbers, so they are purely damped oscillations,
but they have a low real frequency at high wavenumbers. Here and in the remaining part of the paper we consider only Alfv\'en waves.
The dispersion relation in Eq. (\ref{dispn1}) has a very involved analytical solution, therefore it has been solved numerically and the solution is plotted in Fig. 1a.
This plot shows that there is no cut-off wavenumber and it is
similar to the upper plot of Fig. 2 from Zaqarashvili et al.
(2011a).

\begin{figure*}
\begin{center}
\includegraphics[width=5cm, height=4cm]{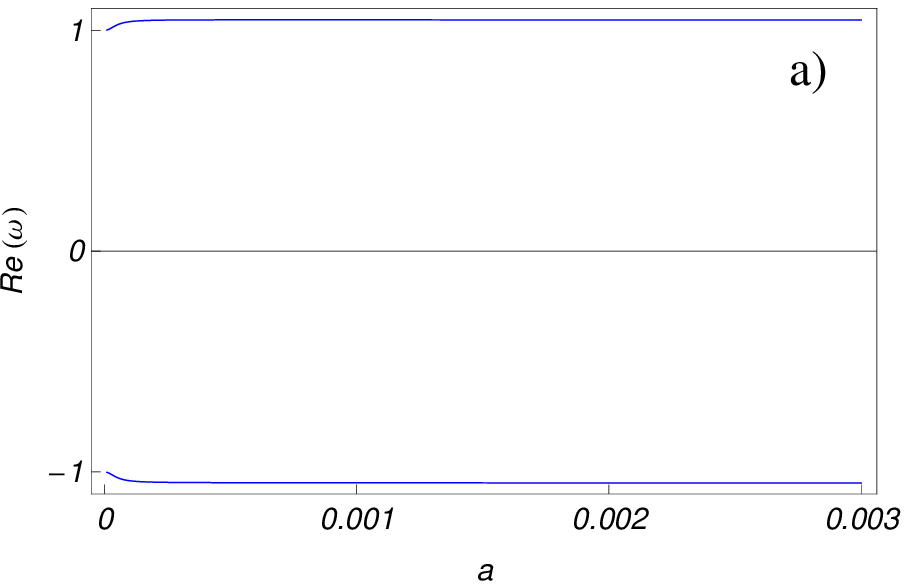}
\includegraphics[width=5cm, height=4cm]{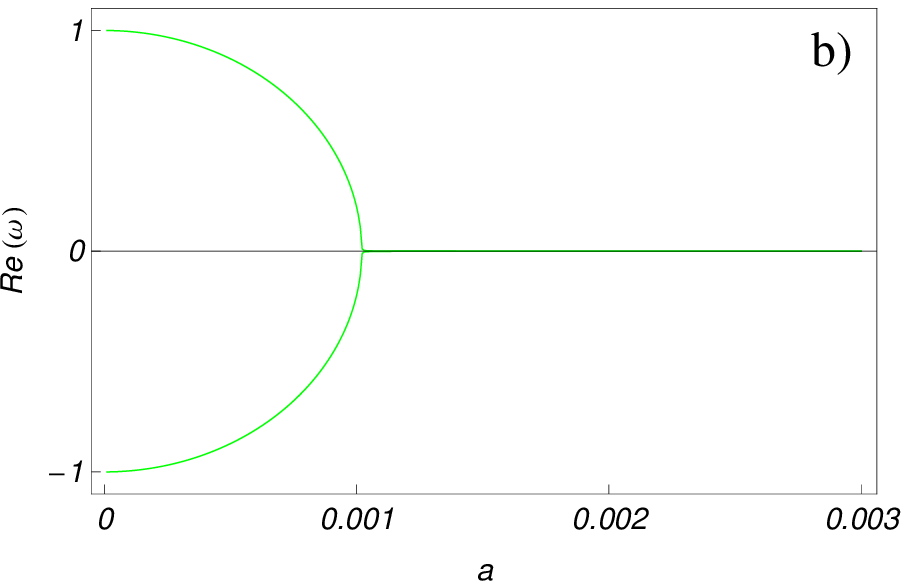}
\end{center}
\end{figure*}

\begin{figure*}
\begin{center}
\includegraphics[width=5cm, height=4cm]{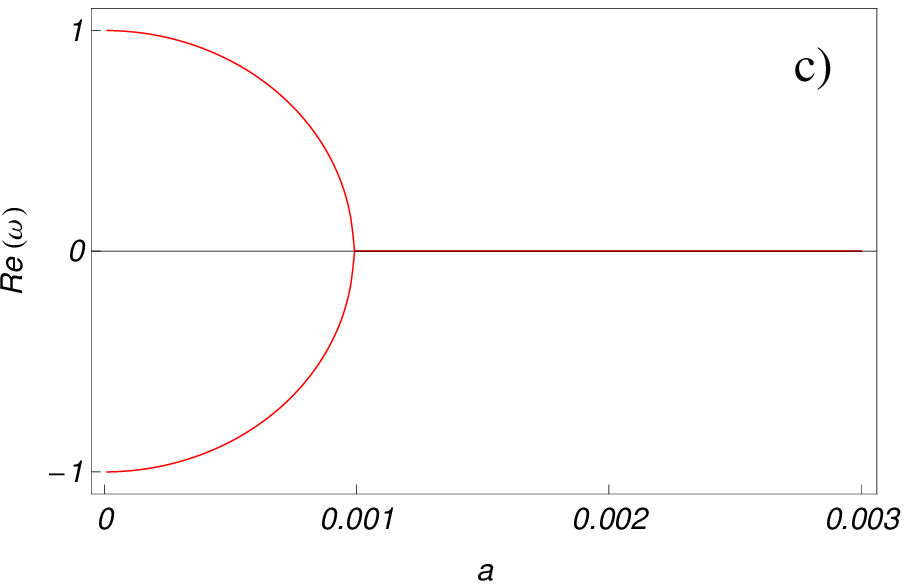}
\includegraphics[width=5cm, height=4cm]{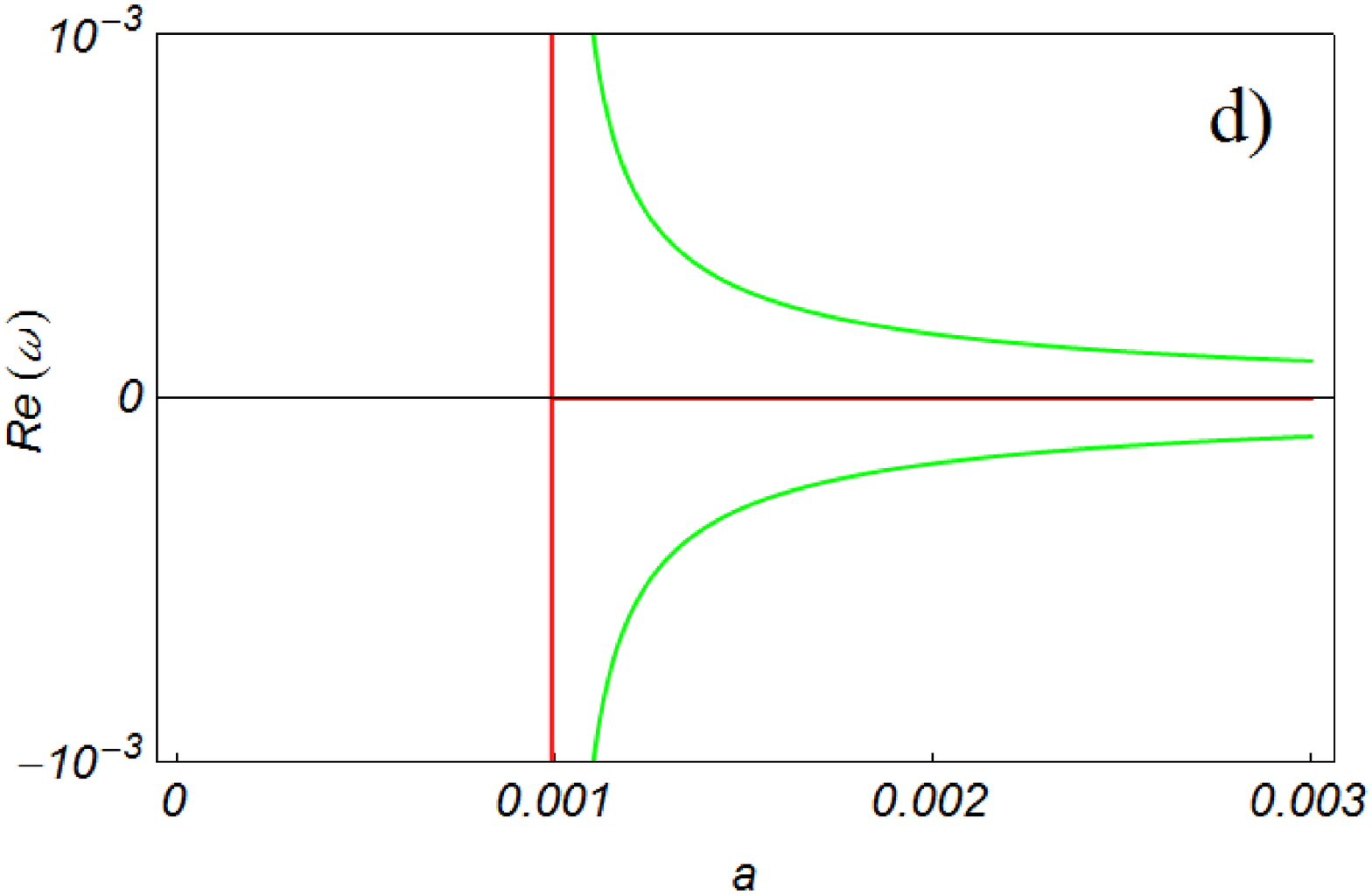}
\end{center}
\caption{Real part of the dimensionless wave frequency $Re(\omega)$ versus
the dimensionless Alfv\'en frequency $a={{k V_A}/ {\omega_i}}$ in
partially ionized plasmas, where ${\omega_i}$ is the ion gyro-frequency.
a) single-fluid MHD equations with inertial term (dispersion relation given by Eq. \ref{dispn1}); b)
single-fluid Hall MHD equations
 (dispersion relation given by Eq. \ref{dispn21});
c) single-fluid MHD equations without modified Hall current (dispersion relation given by Eq. \ref{dispn3}); d) Zoom
of the solutions  in panels b) (green line) and c (red line) near $x$ axis.}
\end{figure*}

The next approximation is neglecting the inertial term in Eq.
(\ref{w1}). This is equivalent to the traditional single-fluid Hall
MHD in partially ionized plasmas, and the dispersion relation is
given by
$$
\delta^2 \nu^2\xi_i^2(1+\zeta)^2\omega^4-2\delta^2 \nu^2 \xi_i^2[1+ \zeta(2+\zeta)]\omega^2-\Big [\delta^4 \nu^2 [1+(1+\nu)\zeta]^2+\xi_n^4+
$$
$$
+2\delta^2 \nu (1+\zeta)\xi_n^2 +\delta^2 \nu^2 \Big (1+\zeta^2(1-4\xi_i)+2\xi_i^2(1+2\zeta)\Big )\Big ]a^2\omega^2 +
$$
$$
+\delta^2 \nu^2 (1+\zeta)^2\xi_i^2+2ia(1+\zeta)\xi_i \delta \nu \left [ \delta \nu (1+(1+\nu)\zeta)+ \xi_n^2 \right ] \omega (\omega^2-1)=
$$
\begin{equation}\label{dispn21}
=0.
\end{equation}
The numerical solution of this dispersion relation is shown in Fig. 1b. At low wavenumbers, the wave frequency is Alfv\'enic, but, when the wavenumber exceeds the value corresponding to the ion-neutral collision frequency ($a\approx 0.00005$ on the $x$ axis), the real part of the wave frequency gradually decreases and becomes very small near $a\approx 0.001$. Since the real part of the frequency never becomes zero, there is no cut-off wavenumber. Thus, the single-fluid approach in partially ionized plasmas with Hall current term does not include a cut-off wavenumber.


Eq. (\ref{dispn21}) can be significantly simplified as $\zeta \ll 1$ i.e. $\alpha_{en}\ll \alpha_{in}$ and it becomes (Pandey and Wardle \cite{Pandey2008})
\begin{equation}\label{dispn2}
\omega^2+\left [i{{\eta_c k}\over V_A}  \pm {{k V_A}\over {\omega_i \xi_i}}\right ] \omega -1=0,
\end{equation}
whose analytical solution is given by
\begin{equation} \label{frehall}
\omega = \frac{1}{2} \left[-\left(\frac{i k \eta_c}{V_A} \pm \frac{k V_A}{\xi_i \omega_i}\right) \pm \sqrt{4+\left(\frac{i k \eta_c}{V_A} \pm \frac{k V_A}{\xi_i \omega_i}\right)^2}\right].
\end{equation}
The real part of the frequency is
\begin{equation} \label{frehall1}
\omega_r = \frac{\sqrt{2}}{4}(A - B)
\end{equation}
with
$$
A = \sqrt{4-\frac{k^2 \eta^2_c}{V^2_A}+\frac{k^2 V^2_A}{\xi_i^2 \omega^2_i}+\sqrt{\frac{4 k^4 \eta^2_c}{\xi_i^2 \omega^2_i}+\left(4+k^2\left(\frac{V^2_A}
{\xi_i^2\omega^2_i} -\frac{\eta^2_c}{V^2_A}\right)\right)^2}}
$$
and
$$
B = \frac{\sqrt{2} k V_A}{\xi_i \omega_i}.
$$
Equations (\ref{frehall}) and (\ref{frehall1}) clearly show that the wave frequency always has a real part, i.e. Hall current term (the second term in front of $\omega$ in Eq. \ref{dispn2}) forbids the appearance of a cut-off
wavenumber.

Next, we neglect the modified Hall current term in Eq. (\ref{B1}),
which gives the dispersion
relation of Alfv\'en waves commonly used in partially ionized
plasmas:
\begin{equation}\label{dispn3}
\omega^2+i{{\eta_c k}\over V_A} \omega -1=0,
\end{equation}
whose analytical solution is
\begin{equation}
\omega = \frac{-i k \eta_c\pm \sqrt{4 V_A^2 - k^2 \eta_C^2}}{2 V_A}. \label{omeganh}
\end{equation}
From Eq.~(\ref{omeganh}), and in dimensional form, the real part of the Alfv\'en frequency can be written as
\begin{equation}
\varpi_r = \pm k \Gamma_A,
\end{equation}
where
\begin{equation}
\Gamma_A = V_A \sqrt{1-\frac{k^2 \eta_C^2}{4 V_A^2}} \label{modalf},
\end{equation}
 represents the modified Alfv\'en speed for the considered case. From this solution, the modified Alfv\'en speed becomes zero when
\begin{equation}\label{dispn4}
k=k_c={{2V_A}\over {\eta_c}},
\end{equation}
and it determines the cut-off wavenumber for Alfv\'en waves. Therefore, Alfv\'en waves cannot propagate for
wavenumbers higher than this cut-off value.



The analytical solution for the real part of the frequency in Eq.(\ref{omeganh}) is plotted in Fig. 1c and becomes zero near $a\approx 0.001$, which
corresponds to the analytical cut-off wavenumber defined by Eq.
(\ref{dispn4}). Furthermore, Fig. 1d displays a zoom near the $x$ axis of the solutions corresponding to dispersion relations given by Eqs. (\ref{dispn21}) and (\ref{dispn3}), showing that only after removing the modified Hall current term in the induction equation the real part of the wave frequency becomes zero.

The condition of evanescence i. e. ($k>k_c$) can be written as
\begin{equation}\label{dispn5}
{{k V_A}\over {\omega_i \xi_i}} \left ( {{\delta_{ei}}\over {\omega_e}} + {{\omega_i}\over {\delta_{in}}} \xi_n \right )>2,
\end{equation}
where
\begin{equation}
\delta_{in}=\alpha_{in}\left ({{1}\over {m_in_i}}+{{1}\over {m_nn_n}}\right )
\end{equation}
is the mean ion-neutral collision frequency, and Eq. (\ref{dispn5}) can be approximated as
 \begin{equation}\label{dispn6}
k V_A>2 \delta_{in}{{\xi_i}\over {\xi_n}},
\end{equation}
which shows that the Alfv\'en frequency corresponding to the cut-off
wavenumber is lower than the mean ion-neutral collision frequency
for weakly ionized plasmas ($\xi_i \ll \xi_n$), and higher for
almost fully ionized plasmas ($\xi_i \gg \xi_n$).

%

Hence, the cut-off wavenumber of Alfv\'en waves in partially ionized plasmas appears after neglecting the inertial term (the left-hand side term in Eq. \ref{w1}) and the modified Hall current term in the induction equation (Eq. \ref{B2}).

Let us estimate the Alfv\'en frequency corresponding to the cut-off wavenumber for parameters typical of solar prominence cores
($n_i$=10$^{10}$ cm$^{-3}$, $n_n$=2${\times}$10$^{10}$ cm$^{-3}$). Then, the cut-off frequency exactly equals the ion-neutral collision frequency 
$\delta_{in}$, which is about 3 s$^{-1}$. For prominence oscillations, the observed periods are in the range between 30 s 
(Balthasar et al. 1993) and 10-30 h (Foullon et al. 2009), therefore, since these periods are much longer than the typical collisional timescales,
the single-fluid approximation can be safely used when interpreting prominence oscillations in terms of MHD waves.

\begin{figure}
		 \includegraphics[width=9cm]{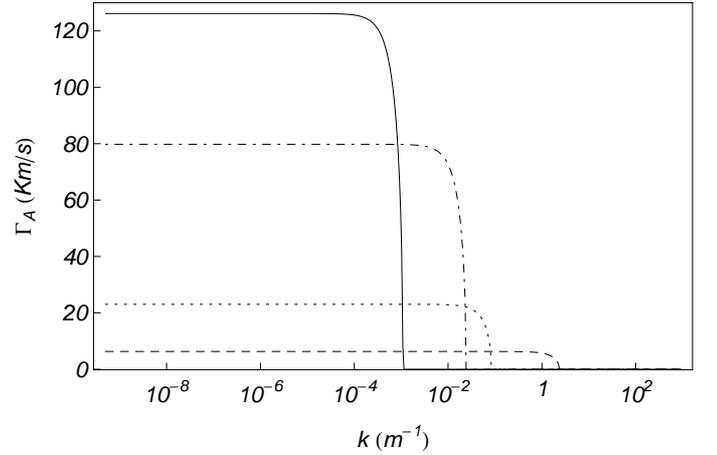}
		   \caption{ Modified Alfv\'en speed, $\Gamma_A$, versus wavenumber for the photosphere (dashed line), the low chromosphere (dotted line), the high chromosphere (dot-dash line), and quiescent prominences (solid line)} \label{layers}
	   \end{figure}

As we stated in Sect.~\ref{intro}, the above calculations are relevant for MHD waves in different layers and structures of the solar atmosphere. Therefore, from the expression for the modified Alfv\'en speed (Eq.~\ref{modalf}), the numerical value for the cut-off wavenumber in different partially ionized layers of the solar atmosphere such as the photosphere and the low and high chromosphere, can be obtained. The physical parameters for these regions of the solar atmosphere are taken from the FAL-1993 model (Fontenla et al. \cite{Fontenla93}), and in Fig.~\ref{layers}  we plot the behavior of the modified Alfv\'en speed, $\Gamma_A$ versus the wavenumber, $k$, for  the layers mentioned above and for quiescent prominences. This figure shows that in all the considered cases a cut-off wavenumber appears, whose numerical value depends, of course, on the different physical conditions that modify  Cowling resistivity. Furthermore, the figure also points out that within a certain range of wavenumbers, the numerical value of the modified Alfv\'en speed for each region or structure considered coincides (horizontal part of the plots) with the ideal Alfv\'en speed, $V_A$, as is expected from Eq.~\ref{modalf}. Finally, we remark that the single-fluid approximation is valid only for waves whose wavenumbers are lower than the cut-off value.

\section{Discussion and conclusion}

Consequent approximations from multi-fluid to single fluid MHD allow
us to find the stage where the cut-off wavenumber for Alfv\'en waves
appears in partially ionized plasmas of the solar atmosphere. The real part of the wave frequency always exists in full
multi-fluid equations, which means that there is no cut-off wavenumber
of Alfv\'en waves. The real part of wave frequency
approaches zero, but never vanishes when the inertia of ion-neutral relative velocity are neglected. This
approximation corresponds to the single-fluid Hall MHD and hence there is no
cut-off wavenumber here, either. On the other hand, the
cut-off wavenumber appears when, after neglecting the inertial term, one neglects the modified Hall current term  in the
induction equation.

The appearance of Alfv\'en waves is due to the current $\vec
j_{\perp}$, which is perpendicular to $\vec b_{\perp}$ and $\vec k$
(see Eq. \ref{B}). This current is due to the ion inertia in an
oscillating electric field $\vec E_{\perp}$ and it plays a key role
in Alfv\'en oscillations through the force $\vec j_{\perp}\times
\vec B$ (Chen \cite{Chen1988}). In the general case, when both
electrons and ions move freely into electric and magnetic fields,
the collision of plasma particles and neutral atoms may lead
only to damping of Alfv\'en waves and waves with any frequency
may propagate in the medium (but remember that the wave frequency
should be always lower than ion-electron collision frequency,
otherwise the fluid approximation is not valid). When one neglects
the electron inertia, i.e. considers the Hall MHD approximation, the electrons are frozen in the magnetic field, but ion inertia is still
at play. In this case, the collision between ions and neutrals becomes
important for shorter scales and it opposes the current $\vec
j_{\perp}$ at higher wavenumbers. However, oscillations again exist
for all wavelengths due to the Hall current, but the wave frequency
is much lower than the Alfv\'en frequency for shorter wavelengths
(see green lines in Fig. 1). The critical point arises when one
neglects the Hall current in the induction equation. Then the
collision between particles completely blocks the current $\vec
j_{\perp}$ at higher wavenumbers and leads to the appearances of
the cut-off frequency.

Finally, there is still no cut-off wavenumber if one neglects the
usual Hall current but retains the electron-neutral collision: the real part of wave frequency very closely approaches the $x$ axis, but never becomes zero. This is because the electrons are not completely frozen into the magnetic field due to the collision with
neutrals, and therefore the ion-neutral collision cannot completely
block the current $\vec j_{\perp}$ at higher wavenumbers.

We note that a cut-off wavenumber may appear for Alfv\'en waves in the stratified solar atmosphere (Murawski, K. \& Musielak, \cite{Murawski2010} and references therein). But this cut-off wavenumber has a completely different origin because it arises as a result of atmospheric stratification due to the solar gravity.

In summary, the cut-off wavenumber in Alfv\'en waves in
single-fluid partially ionized plasmas is due to the approximations made when proceeding from
multifluid to single-fluid equations and is not connected to any
physical process. This conclusion is relevant for MHD wave studies in partially ionized regions and structures of the solar atmos\-phere.

\begin{acknowledgements}
The work was supported by the Austrian Fonds zur F\"orderung
der Wissenschaftlichen Forschung (project P21197-N16) and by the European FP7-PEOPLE-2010-IRSES-269299 project- SOLSPANET. J. L. Ballester and M. Carbonell acknowledge the financial support received from MICINN and FEDER Funds under grant AYA2011-22846 and from CAIB through the "Grups Competitius" scheme. They also acknowledge the warm hospitality enjoyed during the visit payed to the Space Research Institute in Graz where this research was started. J. L. Ballester acknowledges the financial support received from the Austrian Academy of Sciences.
\end{acknowledgements}

\end{document}